\documentclass[
aps,
pra,
reprint,
superscriptaddress,
amsmath,
amssymb
]{revtex4-2}

\usepackage{graphicx}
\usepackage{dcolumn}
\usepackage{bm}
\usepackage{physics}
\usepackage{xcolor}
\usepackage{tikz}
\usepackage[colorlinks=true,citecolor=blue,linkcolor=blue,urlcolor=blue]{hyperref}

\usepackage{amsmath}
\usepackage{amssymb}

\usepackage{ulem}
\usepackage{orcidlink}

\begin{document}

\title{Spin-Selective Spectral Flattening and Wave-Packet Dynamics in a Flux-Engineered Lieb Lattice}

\author{Nana Chang\orcidlink{0000-0003-3631-4426}}
\email{nnchangqq@gmail.com}
\affiliation{State Key Laboratory of Power System Operation and Control, Department of Electrical Engineering, Tsinghua University, Beijing 100089, China}
\author{Xiaoji Zhou\orcidlink{0000-0001-9175-2854}}
\email{xjzhou@pku.edu.cn}
\affiliation{State Key Laboratory of Photonics and Communications, School of Electronics, Peking University, Beijing 100871, China}
\author{Yanglin Zhou\orcidlink{0000-0003-1030-8474}}
\affiliation{State Key Laboratory of Power System Operation and Control, Department of Electrical Engineering, Tsinghua University, Beijing 100089, China}
\author{Song Ci\orcidlink{0000-0002-6535-1974}}
\email{sci@tsinghua.edu.cn}
\affiliation{State Key Laboratory of Power System Operation and Control, Department of Electrical Engineering, Tsinghua University, Beijing 100089, China}

\date{\today}

\title{Spin-Selective Spectral Flattening and Wave-Packet Dynamics in a Flux-Engineered Lieb Lattice}

\begin{abstract}
We investigate reversible internal-state-selective wave-packet
transport induced by spin-dependent Peierls phases in a two-dimensional
nearest-neighbor Lieb lattice. The two conserved spin components
experience effective fluxes
$\alpha_{\sigma}=\alpha_{0}+s_{\sigma}\alpha_{s}$, where
$s_{\uparrow,\downarrow}=\pm1$. At the working point
$\alpha_{0}=\alpha_{s}=1/4$, the spin-up and spin-down components
experience $\alpha_{\uparrow}=1/2$ and $\alpha_{\downarrow}=0$,
respectively. A band-resolved calculation in the $q=2$ magnetic unit
cell shows that the spin-up spectrum contains two zero-energy flat
subbands associated with the sublattice-imbalance flat-band sector,
whereas the remaining four subbands retain finite bandwidths. The
half-flux sector therefore fails the all-bands-flat condition and does
not realize exact Aharonov--Bohm caging for a generic localized initial
state. Nevertheless, real-time simulations reveal a pronounced
suppression of spin-up propagation relative to the dispersive
spin-down component, manifested by a smaller mean-square displacement
and an enhanced finite-region retention probability over the
pre-reflection time window. Reversing the state-dependent flux
interchanges the slow and fast spin channels, while the dynamical
contrast remains robust against moderate flux detuning. These results
establish spin-dependent synthetic flux as a reversible means of
controlling internal-state-resolved matter-wave transport without
spin-flip processes or interactions, and provide complementary
spectral and real-space criteria for distinguishing exact caging from
finite-time dynamical slowing in atomic and photonic flat-band
simulators.
\end{abstract}


\maketitle

\section{Introduction}
\label{sec:introduction}

Controlling motion separately for different internal states is a
central objective of atomic, molecular, and optical quantum simulation\cite{Bloch2008,Bloch2012,Gross2017}.
Such control underlies state-resolved routing, quantum walks, and the
preparation of spatially separated components, but it is difficult to
obtain without introducing spin flips, interactions, or disorder. Flat
bands offer a different route: lattice geometry suppresses kinetic
motion through destructive interference, allowing propagation to be
tuned coherently at the single-particle level\cite{Lieb1989,Leykam2018}.
Foundational flat-band constructions and their modern classification
clarify how lattice connectivity, sublattice imbalance, and compact
localized states constrain this interference
\cite{Mielke1991,Tasaki1992,Rhim2019}.
The two-dimensional Lieb lattice is especially suitable because its
three-site unit cell supports an intrinsic flat band together with two
dispersive bands\cite{Shen2010,Apaja2010,Weeks2010,Goldman2011Lieb}. Localized modes have
been observed in photonic Lieb lattices
\cite{Vicencio2015,Mukherjee2015}, coherent flat-band population has
been realized with ultracold atoms\cite{Taie2015}, and engineered
electronic Lieb lattices have been assembled and spectroscopically
resolved at the atomic scale\cite{Slot2017,Drost2017}. More broadly,
tunable kagome and honeycomb optical lattices demonstrate the geometric
control available in cold-atom platforms\cite{Jo2012,Tarruell2012}.

The remaining challenge is to turn this geometry-induced localization
into a controllable and state-selective transport mechanism. Synthetic
gauge fields provide the required handle: Peierls phases reshape the
interference between hopping paths and reconstruct the spectrum into
magnetic subbands\cite{Hofstadter1976,Dalibard2011,Goldman2014Gauge,
Cooper2019,Goldman2016}. Homogeneous
artificial magnetic fluxes have been implemented through
laser-assisted tunneling in optical lattices
\cite{Spielman2009,Aidelsburger2013,Miyake2013}, enabling the
realization and direct characterization of topological magnetic bands
and their quantum geometry\cite{Jotzu2014,Aidelsburger2015,Atala2013,
Dauphin2013,Price2012,Flaschner2016,Li2016}. Related atomic and photonic
experiments have directly resolved chiral motion and edge transport
under synthetic flux\cite{Atala2014,Mancini2015,Stuhl2015,Hafezi2013}.
At special fluxes, destructive
interference can produce Aharonov--Bohm (AB) cages, for which an
initially localized excitation remains confined to a finite region
\cite{Vidal1998,Vidal2000,Creffield2010,Longhi2014,Mukherjee2018,
Li2022}.

For transport experiments, however, spectral flattening and exact AB
caging are not interchangeable. Narrow magnetic bands can strongly
reduce propagation over a finite observation window even when residual
dispersive channels remain. Conversely, a flat band localizes only the
component of an initial state projected onto its subspace. Establishing
strict caging for a generic localized excitation therefore requires two
complementary tests: all dynamically occupied bands must be flat, and
the real-space propagator must have compact support for all times. This
distinction is essential when localization is inferred from finite-time
density images alone. Quantum-gas microscopy makes such real-space
diagnostics accessible at single-site and single-spin resolution
\cite{Bakr2009,Sherson2010,Weitenberg2011}. It has also enabled direct
measurements of correlation spreading, equilibration, impurity and
spin dynamics, quantum walks, and interacting Hofstadter systems
\cite{Cheneau2012,Trotzky2012,Endres2011,Fukuhara2013,Preiss2015,
Tai2017}.

Internal states make the flux-control problem qualitatively richer.
Light-induced gauge potentials couple internal and motional degrees of
freedom and permit spin-dependent forces and transport
\cite{Lin2011,Beeler2013,Galitski2013,Wu2016,Huang2016,
Goldman2014Gauge}, while state-dependent tunneling
can expose two spin components to opposite or unequal synthetic fluxes
\cite{Aidelsburger2013,Kennedy2013}. This suggests a direct route to a
spin-selective interferometer: one component can be tuned toward a
strongly flattened magnetic spectrum while the other retains dispersive
propagation. Previous work characterized magnetic-field-induced band
deformation and spin-resolved localization tendencies in a Lieb lattice
\cite{chang2026magnetic}. What remains unresolved is whether the
half-flux sector is an exact cage for a generic initial state and how
the spectral difference between the two internal states appears in
real-time observables accessible to atomic and photonic experiments.

Here we answer these questions for noninteracting spin-$1/2$ particles
in a nearest-neighbor Lieb lattice with
$\alpha_{\sigma}=\alpha_{0}+s_{\sigma}\alpha_{s}$. We choose
$\alpha_{0}=\alpha_{s}=1/4$, so that
$(\alpha_{\uparrow},\alpha_{\downarrow})=(1/2,0)$. The longitudinal
Zeeman term is proportional to the orbital identity and produces only a
rigid spin-dependent energy shift; the transport contrast therefore
originates from the spin-dependent Peierls phases. The complete
half-flux spectrum contains two numerically flat and four dispersive
magnetic subbands, which rules out strict all-state AB caging. This
spectral conclusion is independently confirmed by the continued growth
of the spin-up mean-square displacement and by finite leakage from a
fixed observation region.

Despite the absence of exact caging, the flux produces a large and
directly measurable state-selective response. At the representative
time $\tau=4.5$, the spin-up mean-square displacement is approximately
$44\%$ of the spin-down value, and $91.1\%$ of its probability remains
within radius $3a$, compared with $74.6\%$ for spin down. Reversing the
state-dependent flux exchanges the slow and fast components, while a
detuning scan determines the finite parameter window over which the
contrast survives. The resulting mechanism is thus a reversible
internal-state control of coherent wave-packet transport, accompanied
by experimentally accessible criteria that distinguish exact AB caging
from finite-time dynamical slowing in flat-band quantum simulators.

\section{Model and Observables}
\label{sec:model}

\begin{figure*}[t]
    \centering

    \begin{minipage}[c]{0.43\textwidth}
        \centering

        \resizebox{\linewidth}{!}{%
        \begin{tikzpicture}[scale=1.3]

        \tikzset{
            siteA/.style={
                circle,
                shade,
                ball color=lime!80,
                minimum size=18pt,
                inner sep=0pt,
                text=black,
                font=\bfseries
            },
            siteB/.style={
                circle,
                shade,
                ball color=orange!60,
                minimum size=18pt,
                inner sep=0pt,
                text=black,
                font=\bfseries
            },
            siteC/.style={
                circle,
                shade,
                ball color=magenta!60,
                minimum size=18pt,
                inner sep=0pt,
                text=black,
                font=\bfseries
            }
        }

        \def\xoffset{0.05}
        \def\yoffset{0.05}

        \foreach \x in {0,2,4}{
            \foreach \y in {0,2,4}{

                \node[siteC]
                (C\x\y)
                at (\x+1+\xoffset,\y+1+\yoffset) {C};

                \node[siteA]
                (Aup\x\y)
                at (\x+1+\xoffset,\y+2+\yoffset) {A};

                \ifnum\y>0
                    \node[siteA]
                    (Adown\x\y)
                    at (\x+1+\xoffset,\y+\yoffset) {A};
                \fi

                \ifnum\x>0
                    \node[siteB]
                    (Bleft\x\y)
                    at (\x+\xoffset,\y+1+\yoffset) {B};
                \fi

                \node[siteB]
                (Bright\x\y)
                at (\x+2+\xoffset,\y+1+\yoffset) {B};

                \draw[black!70,line width=1.1pt]
                (C\x\y) -- (Aup\x\y);

                \ifnum\y>0
                    \draw[black!70,line width=1.1pt]
                    (C\x\y) -- (Adown\x\y);
                \fi

                \ifnum\x>0
                    \draw[black!70,line width=1.1pt]
                    (C\x\y) -- (Bleft\x\y);
                \fi

                \draw[black!70,line width=1.1pt]
                (C\x\y) -- (Bright\x\y);
            }
        }

        \foreach \x in {0,2,4}{
            \foreach \y in {0,2,4}{
                \draw[
                    dashed,
                    gray,
                    rounded corners=1.5pt
                ]
                (\x+1-0.35+\xoffset,
                 \y+1-0.35+\yoffset)
                rectangle
                (\x+1+1.35+\xoffset,
                 \y+1+1.35+\yoffset);
            }
        }

        \node[font=\normalsize,gray]
        at (3.6+\xoffset,3.5+\yoffset)
        {$(n,m)$};

        \node[font=\normalsize,gray]
        at (3.7+\xoffset,5.6+\yoffset)
        {$(n,m+1)$};

        \node[font=\normalsize,gray]
        at (3.7+\xoffset,1.5+\yoffset)
        {$(n,m-1)$};

        \node[font=\normalsize,gray]
        at (1.7+\xoffset,3.5+\yoffset)
        {$(n-1,m)$};

        \node[font=\normalsize,gray]
        at (5.7+\xoffset,3.6+\yoffset)
        {$(n+1,m)$};

        \draw[->,thick]
        (-0.2,-0.2) -- (6,-0.2)
        node[right] {$n$};

        \draw[->,thick]
        (-0.2,-0.2) -- (-0.2,6)
        node[above] {$m$};

        \draw[thick,red,->]
        (3.6+\xoffset,3.6+\yoffset)
        arc (180:-180:0.25);

        \node[
            red,
            font=\scriptsize,
            fill=white,
            inner sep=1pt
        ]
        at (4.35+\xoffset,4.18+\yoffset)
        {$\uparrow:\alpha_{\uparrow}=1/2$};

        \node[
            blue,
            font=\scriptsize,
            fill=white,
            inner sep=1pt
        ]
        at (4.35+\xoffset,3.88+\yoffset)
        {$\downarrow:\alpha_{\downarrow}=0$};

        \draw[->,thick,purple]
        (3.86+\xoffset,3.38+\yoffset)
        --
        (3.86+\xoffset,4.38+\yoffset)
        node[right] {$B_z$};

        \draw[->,teal,thick]
        (3.0,3.0) -- (4.5,3.0)
        node[below] {$\mathbf a_1$};

        \draw[->,teal,thick]
        (3.0,3.0) -- (3.0,4.5)
        node[left] {$\mathbf a_2$};

        \node[
            rotate=90,
            black,
            font=\large
        ]
        at (0.2,3)
        {Open boundary};

        \node[
            black,
            font=\large
        ]
        at (3,0.2)
        {Open boundary};

        \node[
            anchor=north west,
            font=\bfseries
        ]
        at (0,6.3) {(a)};

        \end{tikzpicture}%
        }

    \end{minipage}\hspace{0.008\textwidth}%
    \begin{minipage}[c]{0.54\textwidth}
        \centering

        \begin{tikzpicture}

            \node[inner sep=0] (img1) {
                \includegraphics[
                    width=0.88\linewidth
                ]{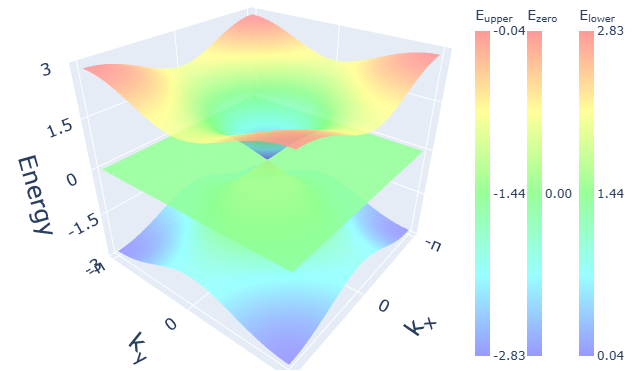}
            };

            \node[
                anchor=north west,
                font=\bfseries,
                fill=white,
                inner sep=1pt
            ]
            at (img1.north west) {(b)};

            \node[
                anchor=north,
                font=\small,
                text=red,
                fill=white,
                inner sep=1pt
            ]
            at (img1.north)
            {$\uparrow:\alpha_{\uparrow}=1/2$};

        \end{tikzpicture}

        \vskip\baselineskip

        \begin{tikzpicture}

            \node[inner sep=0] (img2) {
                \includegraphics[
                    width=0.80\linewidth
                ]{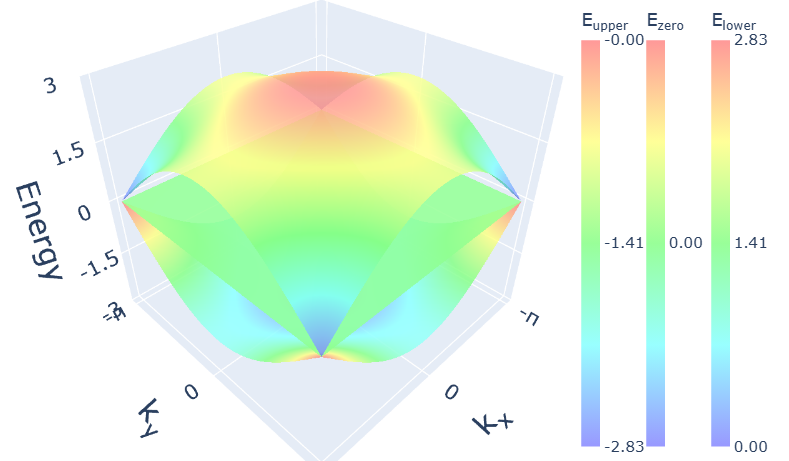}
            };

            \node[
                anchor=north west,
                font=\bfseries,
                fill=white,
                inner sep=1pt
            ]
            at (img2.north west) {(c)};

            \node[
                anchor=north,
                font=\small,
                text=blue,
                fill=white,
                inner sep=1pt
            ]
            at (img2.north)
            {$\downarrow:\alpha_{\downarrow}=0$};

        \end{tikzpicture}

    \end{minipage}

    \caption{
    Spin-dependent synthetic flux and spectral reconstruction in a
    two-dimensional Lieb lattice.
    (a) Schematic of a finite $3\times3$ Lieb-lattice patch containing
    sublattices $A$ (lime), $B$ (orange), and $C$ (magenta).
    Dashed gray squares indicate the central cell $(n,m)$ and its
    neighboring cells $(n\pm1,m)$ and $(n,m\pm1)$.
    Solid lines connect each four-coordinated $C$ site to adjacent
    $A$ and $B$ sites.
    The spin-dependent Peierls phases generate the effective fluxes
    $\alpha_{\uparrow}=1/2$ and $\alpha_{\downarrow}=0$.
    The spin-up component is tuned to half flux, where the magnetic
    subbands are substantially flattened, whereas the spin-down
    component experiences zero synthetic flux.
    The purple arrow denotes an optional longitudinal Zeeman field
    $B_z$, and the teal arrows indicate the primitive lattice vectors
    $\mathbf a_1$ and $\mathbf a_2$.
    Open boundaries are used in the real-time simulations.
   (b) Spin-up magnetic band structure at
    $\alpha_{\uparrow}=1/2$, represented in the doubled magnetic
    unit cell. The flux reconstructs the dispersive sector, while
    two zero-energy flat subbands remain associated with the
    sublattice-imbalance flat-band sector.
    (c) Spin-down band structure at zero flux, retaining dispersive
    Lieb-lattice bands.
    Panels (b) and (c) are calculated using periodic boundary
    conditions; the uniform Zeeman offsets $\pm\Delta_Z$ are omitted
    to expose the orbital dispersions.
    }
    \label{fig:spin_selective_model}

\end{figure*}
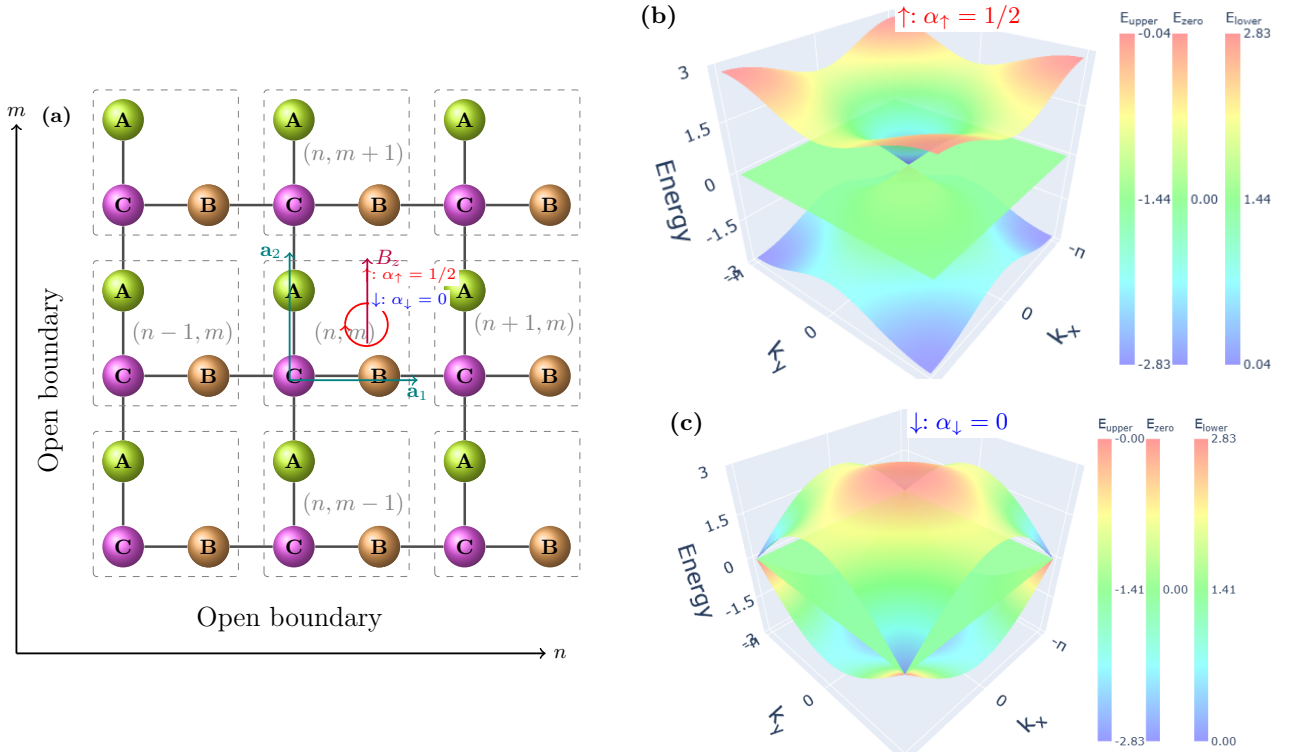

\subsection{Spin-dependent flux on a Lieb lattice}
\label{subsec:Hamiltonian}

We consider noninteracting spin-$1/2$ particles on a
two-dimensional Lieb lattice. Each primitive unit cell contains three
inequivalent sublattice sites, denoted by $A$, $B$, and $C$.
Nearest-neighbor hopping supports an intrinsic flat band originating
from destructive interference between different propagation paths.

A spin-dependent synthetic gauge field is incorporated through the
Peierls substitution,

\begin{equation}
t_{ij}
\longrightarrow
t_{ij}
\exp\left(i\theta_{ij,\sigma}\right),
\end{equation}

where

\begin{equation}
\theta_{ij,\sigma}
=
\frac{2\pi}{\Phi_0}
\int_{\bm r_i}^{\bm r_j}
\bm A_{\sigma}\cdot d\bm l
\end{equation}

is the spin-dependent Peierls phase.
Here, $\Phi_0=h/|Q|$ is the magnetic-flux quantum for a particle
with charge $Q$. For neutral atoms, $\theta_{ij,\sigma}$ represents
a laser-engineered tunneling phase rather than the phase generated
by a physical electromagnetic field.

We decompose the link phase as

\begin{equation}
\theta_{ij,\sigma}
=
\theta_{ij}^{(0)}
+
s_{\sigma}\theta_{ij}^{(s)},
\qquad
s_{\uparrow}=+1,
\quad
s_{\downarrow}=-1,
\label{eq:spin_phase}
\end{equation}

where $\theta_{ij}^{(0)}$ is the spin-independent contribution and
$\theta_{ij}^{(s)}$ is generated by a spin-dependent synthetic gauge
field. The corresponding effective flux experienced by spin
$\sigma$ is

\begin{equation}
\alpha_{\sigma}
\equiv
\frac{\Phi_{\sigma}}{\Phi_0}
=
\alpha_0+s_{\sigma}\alpha_s.
\label{eq:spin_flux}
\end{equation}

The phase accumulated around an elementary plaquette satisfies

\begin{equation}
\sum_{\langle ij\rangle\in\partial p}
\theta_{ij,\sigma}
=
2\pi\alpha_{\sigma}
\pmod{2\pi},
\label{eq:plaquette_phase}
\end{equation}
Here the bonds $\langle ij\rangle\in\partial p$ are oriented
counterclockwise around an elementary plaquette $p$.

Using the spin-dependent Landau gauge

\begin{equation}
\bm A_{\sigma}
=
\left(-B_{\sigma}y,0,0\right),
\end{equation}

the tight-binding Hamiltonian is

\begin{equation}
H
=
-\sum_{\langle i,j\rangle,\sigma}
t_{ij}
\left(
e^{i\theta_{ij,\sigma}}
c_{i\sigma}^{\dagger}c_{j\sigma}
+
\mathrm{H.c.}
\right)
+
\Delta_Z
\sum_i
\left(
c_{i\uparrow}^{\dagger}c_{i\uparrow}
-
c_{i\downarrow}^{\dagger}c_{i\downarrow}
\right),
\label{eq:H}
\end{equation}

where $c_{i\sigma}^{\dagger}$ and $c_{i\sigma}$ are the creation
and annihilation operators for spin $\sigma$ at lattice site $i$.
The Zeeman energy is defined as

\begin{equation}
\Delta_Z
=
\frac{g\mu_BB_z}{2}.
\end{equation}

Because the Zeeman term is proportional to the identity in orbital
space, it produces a rigid spin-dependent energy shift but does not,
by itself, modify the spatial eigenstates or wave-packet spreading.
The spin-selective dynamics studied below therefore originates from
the spin-dependent Peierls phases in Eq.\eqref{eq:spin_phase}.

For a rational effective flux,

\begin{equation}
\alpha_{\sigma}
=
\frac{p_{\sigma}}{q_{\sigma}},
\qquad
p_{\sigma},q_{\sigma}\in\mathbb{Z},
\end{equation}

the magnetic unit cell is enlarged by a factor $q_{\sigma}$.
The magnetic Bloch Hamiltonian for spin $\sigma$ consequently has
dimension $3q_{\sigma}\times3q_{\sigma}$ and is defined over the
corresponding magnetic Brillouin zone.

Diagonalization gives

\begin{equation}
H_{\sigma}(\bm k)
\ket{u_{n\sigma}(\bm k)}
=
E_{n\sigma}(\bm k)
\ket{u_{n\sigma}(\bm k)}.
\label{eq:eigenproblem}
\end{equation}

To maximize the spin-resolved spectral contrast, we choose the gauge
fields such that

\begin{equation}
\alpha_{\uparrow}
=
\frac{1}{2},
\qquad
\alpha_{\downarrow}
\neq
\frac{1}{2}.
\label{eq:working_flux_condition}
\end{equation}

The spin-up component is thereby tuned to half flux, whereas the
spin-down component retains a different effective flux and a larger
dispersive bandwidth. Whether the half-flux sector is strictly caged
must be determined from the complete set of magnetic subband widths
and from real-time propagation; the plaquette phase alone is not a
sufficient criterion.

Figure\ref{fig:spin_selective_model} summarizes the lattice geometry
and the mechanism of spin-selective spectral reconstruction.
Panel (a) shows the finite Lieb lattice used for the real-time
simulations. The four-coordinated $C$ sites are connected to the
vertical-edge $A$ sites and horizontal-edge $B$ sites, and the cells
are indexed by $(n,m)$. The primitive vectors $\mathbf a_1$ and
$\mathbf a_2$ define the two lattice directions. Open boundaries are
used in this panel so that the spatial evolution of an initially
localized wave packet can be monitored directly.

For the representative configuration displayed in
Fig.\ref{fig:spin_selective_model}, we take
$\alpha_0=\alpha_s=1/4$, giving
$\alpha_{\uparrow}=1/2$ and $\alpha_{\downarrow}=0$. The spin-up
component therefore accumulates a plaquette phase
$2\pi\alpha_{\uparrow}=\pi$. This phase reconstructs the magnetic
spectrum and partially flattens its magnetic subbands, as
illustrated qualitatively in Fig.\ref{fig:spin_selective_model}(b)
and quantified in Fig.\ref{fig:spin_bands}. By contrast, the
spin-down component experiences zero synthetic flux and retains the
intrinsic zero-energy flat band together with two dispersive
Lieb-lattice bands [Fig.\ref{fig:spin_selective_model}(c)]. The
resulting wave-packet dynamics depend on the overlap of the initial
state with both the flat and dispersive subspaces.

Panels (b) and (c) are Bloch spectra calculated with periodic boundary
conditions, whereas panel (a) represents the finite open-boundary
lattice used for time evolution. The uniform Zeeman term shifts the
two spectra rigidly by $\pm\Delta_Z$ but does not alter their orbital
eigenvectors or bandwidths. The offsets are therefore removed in
panels (b) and (c). Any spin-dependent difference in orbital dynamics
therefore originates from the synthetic flux rather than from the
Zeeman splitting.

\subsection{Density of states and spectral diagnostics}
\label{subsec:band_geometry}

The spin-resolved density of states is calculated as

\begin{equation}
\rho_{\sigma}(E)
=
\frac{1}{q_{\sigma}N_k}
\sum_{n,\bm k}
\delta
\left[
E-E_{n\sigma}(\bm k)
\right],
\label{eq:DOS}
\end{equation}

where $N_k$ is the number of sampled momenta and the factor
$q_{\sigma}$ normalizes the result per primitive unit cell.
Numerically, the delta function is approximated by Gaussian
broadening,

\begin{equation}
\delta(E-E_{n\sigma})
\longrightarrow
\frac{1}{\sqrt{2\pi}\eta}
\exp\left[
-\frac{(E-E_{n\sigma})^2}{2\eta^2}
\right],
\end{equation}

where $\eta$ is the broadening parameter. The total density of states
is

\begin{equation}
\rho(E)
=
\rho_{\uparrow}(E)
+
\rho_{\downarrow}(E).
\end{equation}

The bandwidth of band $n$ is defined as

\begin{equation}
W_{n\sigma}
=
\max_{\bm k}E_{n\sigma}(\bm k)
-
\min_{\bm k}E_{n\sigma}(\bm k).
\label{eq:bandwidth}
\end{equation}

An all-bands-flat spectrum requires $W_{n\sigma}=0$ for every magnetic
subband that overlaps the initial state. This provides a direct and
model-independent spectral test of strict caging. If only a subset of
the bands is flat, the dynamics are generally state dependent and may
display partial localization or a long finite-time plateau rather than
compact confinement.

Energies are expressed in units of the nearest-neighbor hopping $t$.
The band structures in Fig.\ref{fig:spin_bands}(b) and (c) use 150
points per high-symmetry segment. The working-point bandwidths are
converged on a $51\times51$ magnetic-momentum mesh. The density of
states uses a $91\times91$ mesh and Gaussian broadening
$\eta=0.045t$. Band-resolved widths and densities of states are
recomputed in the appropriate magnetic Brillouin zone for each spin
sector. The detuning analysis below instead uses a unit-cell-independent
total spectral width, because an average over a flux-dependent number
of folded magnetic subbands is not directly comparable across arbitrary
rational fluxes.

\subsection{Spin-resolved real-time dynamics}
\label{subsec:dynamics}

To investigate spin-selective transport and localization, we consider
an initially localized wave packet,

\begin{equation}
\psi_{i\sigma}(0)
=
\frac{a_{\sigma}}{\sqrt{\mathcal N}}
\exp\left[
-\frac{
|\bm r_i-\bm r_0|^2
}{
2w^2
}
\right]
\chi_i,
\label{eq:initial_state}
\end{equation}

where $\bm r_0$ is the initial center, $w$ is the spatial width,
$\chi_i$ specifies the initially occupied sublattice, and
$a_{\sigma}$ determines the initial spin composition. The
normalization factor $\mathcal N$ is chosen such that

\begin{equation}
\sum_{i,\sigma}
|\psi_{i\sigma}(0)|^2
=
1.
\end{equation}

Unless otherwise specified, we use an equal coherent spin
superposition,

\begin{equation}
a_{\uparrow}
=
a_{\downarrow}
=
\frac{1}{\sqrt{2}}.
\end{equation}

The state in each conserved spin sector evolves according to

\begin{equation}
\ket{\psi_{\sigma}(t)}
=
\exp\left(
-\frac{i}{\hbar}H_{\sigma}t
\right)
\ket{\psi_{\sigma}(0)}.
\label{eq:evolution}
\end{equation}

The spin-resolved local density is

\begin{equation}
n_{i\sigma}(t)
=
\mel{\psi_{\sigma}(t)}
{c_{i\sigma}^{\dagger}c_{i\sigma}}
{\psi_{\sigma}(t)}.
\label{eq:density}
\end{equation}

The corresponding mean-square displacement is

\begin{equation}
\mathrm{MSD}_{\sigma}(t)
=
\frac{
\displaystyle
\sum_i
|\bm r_i-\bm r_0|^2
n_{i\sigma}(t)
}{
\displaystyle
\sum_i n_{i\sigma}(t)
}.
\label{eq:MSD}
\end{equation}

The time-dependent effective spreading exponent is defined by

\begin{equation}
\alpha_{\sigma}^{\mathrm{eff}}(t)
=
\frac{
d\ln\left[\mathrm{MSD}_{\sigma}(t)\right]
}{
d\ln t
}.
\label{eq:transport_exponent}
\end{equation}

Ballistic propagation is characterized by
$\alpha_{\sigma}^{\mathrm{eff}}\simeq2$, whereas saturation of the
MSD gives $\alpha_{\sigma}^{\mathrm{eff}}\rightarrow0$. In the
present clean and closed system, an intermediate exponent should not
be identified as diffusive unless an explicit scattering or
dephasing mechanism is introduced.

The spin-resolved participation ratio is

\begin{equation}
\mathrm{PR}_{\sigma}(t)
=
\frac{
\left[
\sum_i n_{i\sigma}(t)
\right]^2
}{
\sum_i n_{i\sigma}^2(t)
}.
\label{eq:PR}
\end{equation}

A bounded participation ratio indicates confinement to a finite
number of sites, whereas a growing participation ratio signals
spatial spreading.

We also introduce a finite-region retention probability,

\begin{equation}
P_{\mathcal R,\sigma}(t)
=
\frac{
\displaystyle
\sum_{i\in\mathcal R}
n_{i\sigma}(t)
}{
\displaystyle
\sum_i n_{i\sigma}(t)
},
\label{eq:retention_probability}
\end{equation}

where $\mathcal R$ is a fixed cluster centered on the initial wave
packet. Unlike a cage-survival probability, this observable does not
presuppose exact compact confinement: its decay directly measures
leakage from the observation region.

Finally, the dynamical separation between the two spin components is
quantified by the MSD contrast

\begin{equation}
\mathcal C_{\mathrm{MSD}}(t)
=
\frac{
\mathrm{MSD}_{\downarrow}(t)
-
\mathrm{MSD}_{\uparrow}(t)
}{
\mathrm{MSD}_{\downarrow}(t)
+
\mathrm{MSD}_{\uparrow}(t)
}.
\label{eq:MSD_contrast}
\end{equation}

A positive $\mathcal C_{\mathrm{MSD}}(t)$ indicates that the spin-down
component has spread farther than the spin-up component. In a finite
system it is evaluated only before boundary reflections become
important; no asymptotic value is imposed a priori.

The time evolution is calculated by exact diagonalization on finite
Lieb lattices with open boundary conditions. The maximum propagation
time is restricted to times shorter than the boundary-return time.
Convergence is verified with respect to lattice size, momentum-grid
resolution, Gaussian broadening, and time discretization.

\section{Results}
\label{sec:results}

\subsection{Spin-selective spectral reconstruction}
\label{subsec:band_reconstruction}

\begin{figure*}[t]
    \centering
    \includegraphics[width=\textwidth]
    {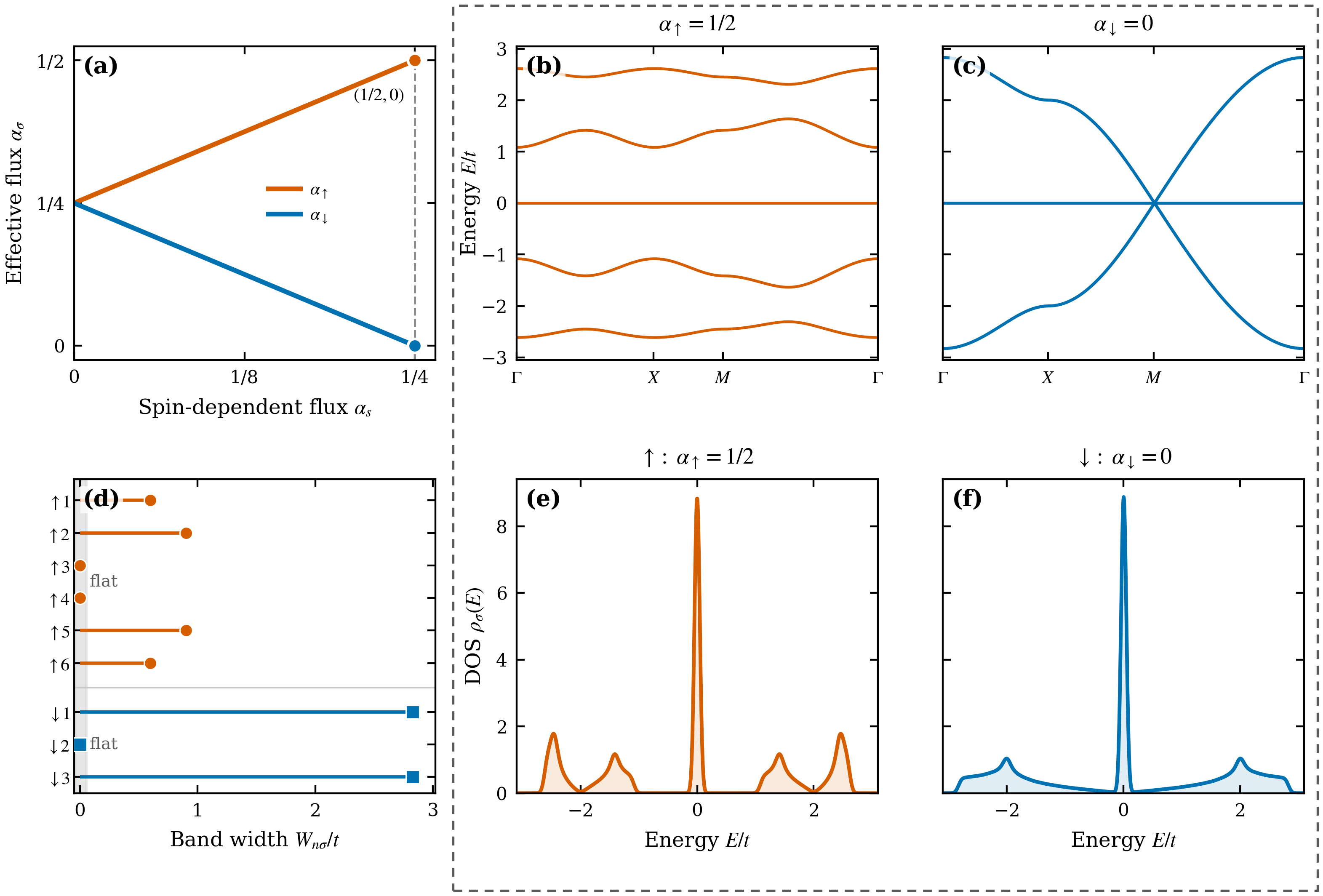}
    \caption{
    Spin-selective reconstruction of the Lieb-lattice spectrum.
    (a) Effective fluxes
    $\alpha_{\uparrow}=\alpha_{0}+\alpha_{s}$ and
    $\alpha_{\downarrow}=\alpha_{0}-\alpha_{s}$ as functions of the
    spin-dependent flux $\alpha_{s}$ for $\alpha_{0}=1/4$.
    The marked working point $\alpha_{s}=1/4$ gives
    $(\alpha_{\uparrow},\alpha_{\downarrow})=(1/2,0)$.
    (b) Spin-up magnetic bands at $\alpha_{\uparrow}=1/2$.
    Two of the six magnetic subbands are dispersionless within numerical
    precision, while the remaining bands retain finite bandwidths.
    (c) Spin-down bands at $\alpha_{\downarrow}=0$, consisting of one
    intrinsic flat band and two dispersive bands.
    (d) Band-resolved bandwidths $W_{n\sigma}$ at the working point;
    the shaded strip identifies numerically flat bands.
    (e) Spin-up and (f) spin-down densities of states.
    The dashed rectangle groups the spin-resolved band structures with
    their corresponding densities of states.
    Orange and blue denote the spin-up and spin-down sectors,
    respectively.
    A spin-independent energy offset and the longitudinal Zeeman shift
    have been removed from each spectrum to expose the orbital
    reconstruction.
    }
    \label{fig:spin_bands}
\end{figure*}

We first examine how the spin-dependent synthetic flux reconstructs
the Lieb-lattice spectrum. The effective fluxes experienced by the two
spin components are

\begin{equation}
\alpha_{\uparrow}=\alpha_{0}+\alpha_{s},
\qquad
\alpha_{\downarrow}=\alpha_{0}-\alpha_{s},
\label{eq:effective_fluxes}
\end{equation}

where $\alpha_{0}$ is the spin-independent background flux and
$\alpha_{s}$ denotes the spin-dependent contribution. Throughout the
following analysis, we choose

\begin{equation}
\alpha_{0}=\alpha_{s}=\frac{1}{4},
\label{eq:working_point}
\end{equation}

which realizes the spin-selective working point

\begin{equation}
\left(
\alpha_{\uparrow},
\alpha_{\downarrow}
\right)
=
\left(
\frac{1}{2},0
\right).
\label{eq:spin_selective_flux_point}
\end{equation}

As illustrated in Fig.\ref{fig:spin_bands}(a), the two spin
components therefore experience markedly different orbital fluxes
despite occupying the same lattice. Because the Hamiltonian is
diagonal in spin, the two sectors can be analyzed independently.
The longitudinal Zeeman term produces only a rigid spin-dependent
energy displacement and is subtracted in Fig.\ref{fig:spin_bands}
so that the flux-induced reconstruction of the orbital spectrum can
be compared directly.

At $\alpha_{\uparrow}=1/2$, the magnetic unit cell is doubled and the
three zero-field Lieb bands split into six magnetic subbands.
Figure\ref{fig:spin_bands}(b) shows that two of these subbands become
flat within numerical precision. The other four subbands, however,
remain dispersive. Their numerically determined bandwidths are

\begin{equation}
\begin{split}
\frac{1}{t}
\left(
W_{1\uparrow},
W_{2\uparrow},
W_{3\uparrow},
W_{4\uparrow},
W_{5\uparrow},
W_{6\uparrow}
\right)\\
=
\left(
0.598,\,
0.902,\,
0,\,
0,\,
0.902,\,
0.598
\right).
\end{split}
\label{eq:up_bandwidths}
\end{equation}

Thus, half flux produces a substantial spectral flattening but not a
complete collapse of the spin-up spectrum.

By contrast, the spin-down sector experiences zero effective flux.
Its spectrum retains the conventional three-band structure of the
nearest-neighbor Lieb lattice, comprising a central flat band and two
strongly dispersive bands, as shown in
Fig.\ref{fig:spin_bands}(c). The corresponding bandwidths are

\begin{equation}
\frac{1}{t}
\left(
W_{1\downarrow},
W_{2\downarrow},
W_{3\downarrow}
\right)
=
\left(
2.827,\,
0,\,
2.827
\right).
\label{eq:down_bandwidths}
\end{equation}

For a quantitative comparison, we define the mean band-resolved
bandwidth in each spin sector as

\begin{equation}
\overline{W}_{\sigma}
=
\frac{1}{N_{\sigma}}
\sum_{n=1}^{N_{\sigma}}W_{n\sigma},
\label{eq:mean_bandwidth}
\end{equation}

where $N_{\uparrow}=6$ and $N_{\downarrow}=3$ at the selected working
point. The calculated values are

\begin{equation}
\overline{W}_{\uparrow}=0.500t,
\qquad
\overline{W}_{\downarrow}=1.885t.
\label{eq:mean_bandwidth_values}
\end{equation}

Within their natural magnetic representations, the spin-up subbands
are therefore narrower on average than the spin-down bands, although
the spin-up spectrum is not completely flat. Because
$N_{\uparrow}\neq N_{\downarrow}$, however,
$\overline{W}_{\sigma}$ depends on magnetic-band folding and is not
invariant under an arbitrary enlargement of the unit cell. It is used
here only as a descriptive compression measure. The physical comparison
is based on the complete band-resolved widths in
Fig.\ref{fig:spin_bands}(d) and on real-time propagation from identical
initial states.

The corresponding densities of states are presented in
Figs.\ref{fig:spin_bands}(e) and \ref{fig:spin_bands}(f).
The sharp peaks originate from the flat or nearly flat magnetic
subbands, whereas the broader structures reflect the residual
dispersive spectrum. In particular, the half-flux spin-up sector
exhibits a redistribution of spectral weight associated with magnetic
band reconstruction. The zero-flux spin-down sector retains the
characteristic flat-band peak at $E=0$ together with the continuum
generated by the two dispersive bands.

These results establish a strong spin-dependent contrast in the
available propagation channels. They do not, by themselves, establish
strict Aharonov--Bohm caging: because four spin-up magnetic subbands
retain finite bandwidths, a generic localized state can still overlap
with propagating modes. We therefore next distinguish exact caging
from localization restricted to the flat-band subspace.

\subsection{Testing the Aharonov--Bohm-caging criterion}
\label{subsec:caging_test}

We next examine whether the half-flux spin sector satisfies the
conditions for Aharonov--Bohm caging. For a generic localized initial
state, strict caging requires that every dynamically occupied magnetic
band be dispersionless,

\begin{equation}
W_{n\sigma}=0
\qquad
\text{for all occupied }n,
\label{eq:all_flat_criterion}
\end{equation}

and that the corresponding real-space propagator have compact support.
Specifically, there must exist a finite region $\mathcal R_j$
surrounding the initially populated site $j$ such that

\begin{equation}
G_{ij,\sigma}(t)
=
\mel{i}
{
e^{-iH_\sigma t/\hbar}
}
{j}
=
0,
\qquad
i\notin\mathcal R_j,
\label{eq:compact_propagator_criterion}
\end{equation}

for all evolution times. Equations\eqref{eq:all_flat_criterion} and
\eqref{eq:compact_propagator_criterion} distinguish exact caging from
a transient reduction of the propagation velocity.

The spectrum in Fig.\ref{fig:spin_bands}(b) does not satisfy the
all-bands-flat criterion. At $\alpha_{\uparrow}=1/2$, two of the six
magnetic subbands are flat within numerical precision, whereas the
remaining four retain the finite bandwidths listed in
Eq.\eqref{eq:up_bandwidths}. Consequently, a generic localized
spin-up wave packet contains dispersive components that can propagate
outside any finite region.

To separate the flat and dispersive contributions, we introduce the
projector onto the spin-resolved flat-band subspace,

\begin{equation}
\hat{\mathcal P}_{\mathrm{flat},\sigma}
=
\sum_{n\in\mathcal F_\sigma}
\sum_{\bm k}
\ket{u_{n\sigma}(\bm k)}
\bra{u_{n\sigma}(\bm k)},
\label{eq:flat_projector}
\end{equation}

where $\mathcal F_\sigma$ denotes the set of numerically flat magnetic
bands. The flat-subspace weight of the initial state is

\begin{equation}
\mathcal W_{\mathrm{flat},\sigma}
=
\mel{\psi_\sigma(0)}
{
\hat{\mathcal P}_{\mathrm{flat},\sigma}
}
{\psi_\sigma(0)}.
\label{eq:flat_weight}
\end{equation}

The time-evolved state can then be decomposed as

\begin{align}
\ket{\psi_\sigma(t)}
={}&
e^{-iH_\sigma t/\hbar}
\hat{\mathcal P}_{\mathrm{flat},\sigma}
\ket{\psi_\sigma(0)}
\nonumber\\
&+
e^{-iH_\sigma t/\hbar}
\left(
1-\hat{\mathcal P}_{\mathrm{flat},\sigma}
\right)
\ket{\psi_\sigma(0)}.
\label{eq:flat_dispersive_decomposition}
\end{align}

The first term is nondispersive, whereas the second evolves through the
residual dispersive magnetic subbands. If the projected component is
prepared as a compact localized state, it remains confined within the
flat-band subspace. By contrast, a single-site or Gaussian initial state
generally has $\mathcal W_{\mathrm{flat},\uparrow}<1$ and can therefore
exhibit finite leakage.

The localization behavior is also sublattice dependent. At zero flux,
the intrinsic nearest-neighbor Lieb flat band has vanishing amplitude
on the four-coordinated hub sublattice. A hub-site excitation therefore
has no overlap with that intrinsic flat band, whereas an appropriately
phased superposition on the edge sublattices can selectively populate a
compact localized state. At finite flux, the corresponding overlap must
be evaluated for the reconstructed magnetic eigenstates. The initially
occupied sublattice and relative phases must therefore be specified when
assessing localization.

Accordingly, the model permits flat-subspace localization for suitably
prepared states and produces a pronounced suppression of spin-up
propagation for the central-site state considered below, but it does not
realize strict all-state Aharonov--Bohm caging. The real-time observables
determine how this distinction appears in experimentally accessible
dynamics.

\subsection{Real-time dynamical signatures}
\label{subsec:real_time_dynamics}
\begin{figure*}[t]
    \centering
    \includegraphics[width=0.96\textwidth]
    {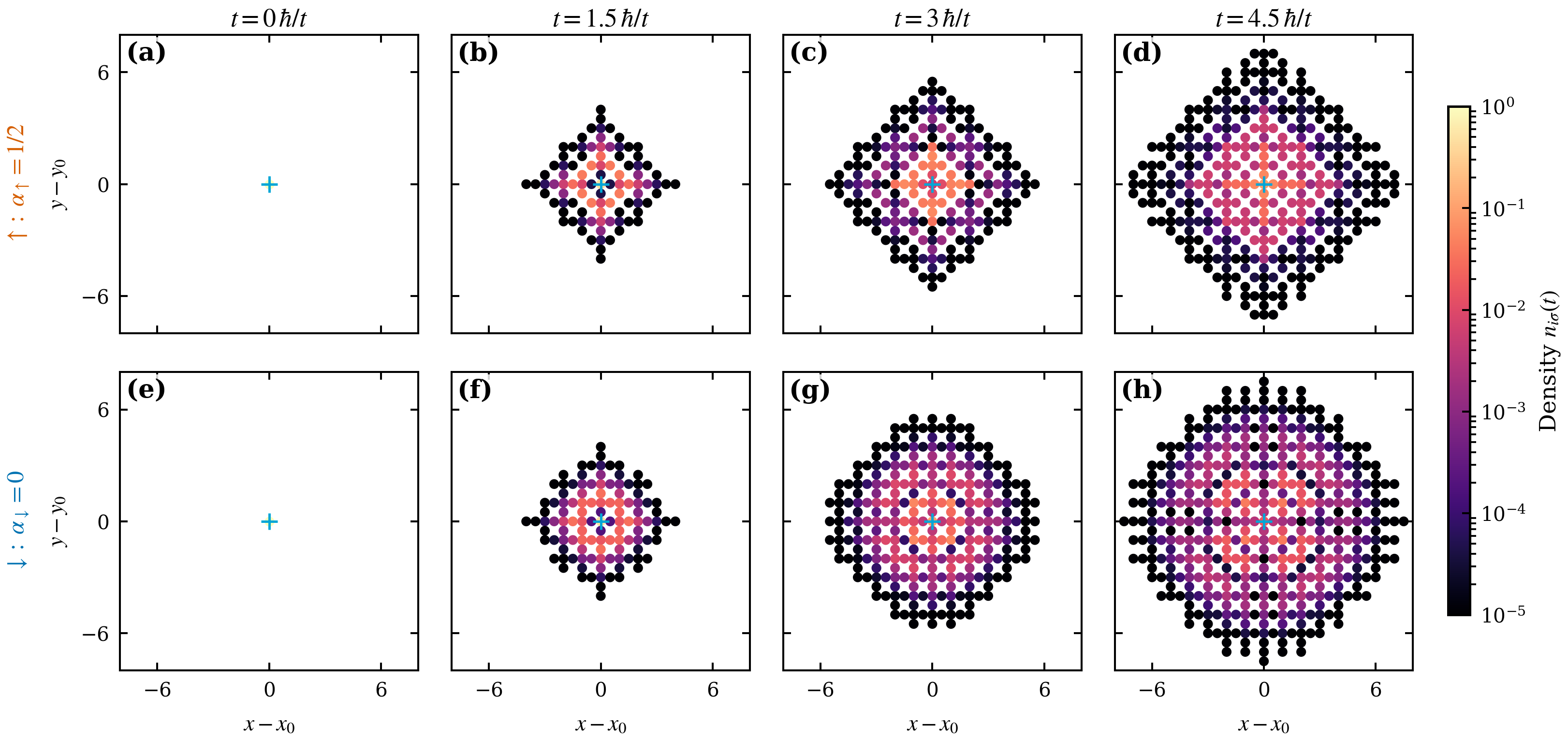}
    \caption{
    Spin-resolved real-time wave-packet dynamics. The upper and lower
    rows show the spin-up and spin-down density distributions,
    respectively, at the indicated dimensionless evolution times
    $\tau=t_{\mathrm{hop}}t/\hbar$. At
    $\alpha_{\uparrow}=1/2$, the spin-up component displays slower
    spreading and enhanced central retention, whereas the spin-down
    component spreads more rapidly at $\alpha_{\downarrow}=0$.
    All density panels use the same logarithmic color scale.
    The cyan cross marks the initially populated central $C$ site.
    }
    \label{fig:density_dynamics}
\end{figure*}

We now examine how the spin-dependent spectral reconstruction appears
in the real-time wave-packet dynamics. The two spin components are
prepared with identical spatial profiles and equal initial populations.
Their subsequent dynamical difference therefore originates exclusively
from the spin-dependent Peierls phases.

To avoid confusion between the physical time and the hopping energy,
we introduce the dimensionless time

\begin{equation}
\tau
=
\frac{t_{\mathrm{hop}}t}{\hbar},
\label{eq:dimensionless_time}
\end{equation}

where $t_{\mathrm{hop}}$ denotes the nearest-neighbor hopping energy.
For the calculations shown in Fig.\ref{fig:density_dynamics}, the
initial state is localized on the central $C$ site,

\begin{equation}
\ket{\psi_{\sigma}(0)}
=
\ket{C,\boldsymbol{r}_0}.
\label{eq:figure3_initial_state}
\end{equation}

The finite open lattice contains a $25\times25$ array of $C$ sites
and 1825 lattice sites in total.

Figure\ref{fig:density_dynamics} shows the spin-resolved density
distributions at

\begin{equation}
\tau=0,\quad 1.5,\quad 3.0,\quad 4.5.
\end{equation}

At $\tau=0$, the two spin components have identical density profiles.
Their spatial distributions subsequently separate because they
experience different effective magnetic fluxes.

At $\alpha_{\uparrow}=1/2$, the spin-up wave packet develops an
anisotropic interference pattern and expands progressively away from
the initial site. Its spatial extent nevertheless remains smaller than
that of the zero-flux component over the complete observation window.
This reduced spreading is consistent with the reconstructed and
partially flattened spectrum shown in Fig.\ref{fig:spin_bands}(b) and
quantified band by band in Fig.\ref{fig:spin_bands}(d).

The spin-down component at $\alpha_{\downarrow}=0$ develops a broader
propagating front. Its faster expansion reflects the larger group
velocities of the two dispersive zero-flux Lieb bands. At the final
displayed time, $\tau=4.5$, the mean-square displacements are

\begin{equation}
\mathrm{MSD}_{\uparrow}(4.5)
=
3.348\,a^2,
\qquad
\mathrm{MSD}_{\downarrow}(4.5)
=
7.673\,a^2.
\label{eq:figure3_msd}
\end{equation}

The corresponding MSD ratio is

\begin{equation}
R_{\mathrm{MSD}}(4.5)
=
\frac{
\mathrm{MSD}_{\uparrow}(4.5)
}{
\mathrm{MSD}_{\downarrow}(4.5)
}
=
0.436.
\label{eq:figure3_msd_ratio}
\end{equation}

Thus, the spin-up mean-square spatial extent is approximately $44\%$
of the spin-down value at the final displayed time.

The difference is independently quantified using the finite-region
retention probability. Taking $\mathcal R$ to be a circular region of
radius $3a$ centered at $\boldsymbol{r}_0$, we obtain

\begin{equation}
P_{\mathcal R,\uparrow}(4.5)
=
0.911,
\qquad
P_{\mathcal R,\downarrow}(4.5)
=
0.746.
\label{eq:figure3_retention}
\end{equation}

The half-flux component therefore retains a larger fraction of its
density near the initial position. The propagating fronts remain well
separated from the open boundaries throughout the displayed interval,
excluding boundary reflections as the origin of the observed contrast.

Because four spin-up magnetic subbands retain finite bandwidth, the
density in the upper row continues to expand. Figure
\ref{fig:density_dynamics} therefore demonstrates spin-selective
finite-time dynamical slowing and enhanced central retention rather
than strict compact caging.

\subsection{Dynamical slowing and spin contrast}
\label{subsec:dynamical_contrast}

\begin{figure*}[t]
    \centering
    \includegraphics[width=\textwidth]
    {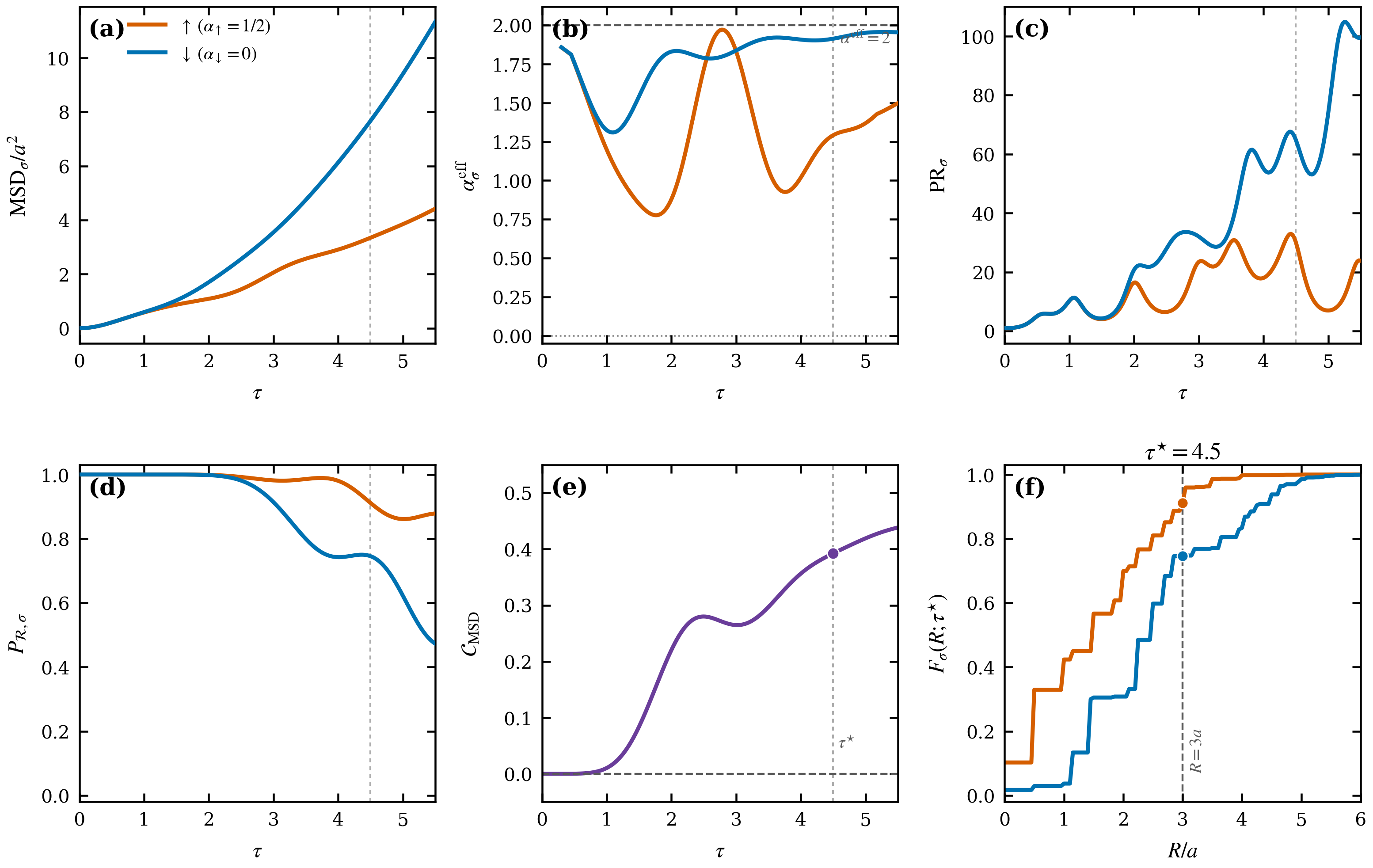}
    \caption{
    Quantitative characterization of spin-selective wave-packet
    dynamics at
    $(\alpha_{\uparrow},\alpha_{\downarrow})=(1/2,0)$.
    (a) Spin-resolved mean-square displacement
    $\mathrm{MSD}_{\sigma}$.
    (b) Local effective spreading exponent
    $\alpha_{\sigma}^{\mathrm{eff}}$; the horizontal dashed line marks
    the ballistic value $\alpha^{\mathrm{eff}}=2$.
    Values at $\tau<0.3$ are omitted because the logarithmic derivative
    is ill-conditioned in the short-time limit.
    (c) Spin-resolved participation ratio
    $\mathrm{PR}_{\sigma}$.
    (d) Probability $P_{\mathcal R,\sigma}$ retained within the region
    $\mathcal R=\{i:|\mathbf r_i-\mathbf r_0|\leq3a\}$.
    (e) Normalized MSD contrast
    $\mathcal C_{\mathrm{MSD}}$.
    The vertical dotted lines in (a)--(e) identify the representative
    observation time $\tau^{\star}=4.5$.
    (f) Cumulative radial probability
    $F_{\sigma}(R;\tau^{\star})$ evaluated at the same time; the vertical
    dashed line marks $R=3a$, and the colored symbols reproduce the
    retention probabilities in (d).
    Orange and blue denote the spin-up and spin-down components,
    respectively, while purple denotes the MSD contrast.
    Solid curves correspond to a lattice of $25\times25$ unit cells.
    The shaded envelopes show the range obtained for
    $L=21$, $25$, and $29$ and represent finite-size variation rather
    than statistical uncertainty.
    The initial state is localized on the central $C$ site, and
    $\tau=t_{\mathrm{hop}}t/\hbar$ is the dimensionless evolution time.
    }
    \label{fig:dynamical_observables}
\end{figure*}

Following the real-space density evolution shown in
Fig.\ref{fig:density_dynamics}, we now quantify the dynamical
difference between the two spin components. The initial state is
localized on the central $C$ site of a finite Lieb lattice with open
boundaries. Unless otherwise stated, the solid curves are calculated
for a system containing $25\times25$ unit cells. Results for
$L=21$, $25$, and $29$ are used to assess the sensitivity to the
finite system size.

Figure\ref{fig:dynamical_observables}(a) shows the spin-resolved
mean-square displacement. The spin-down component at
$\alpha_{\downarrow}=0$ expands rapidly, whereas the half-flux
spin-up component exhibits substantially slower spatial spreading.
At the representative observation time $\tau^{\star}=4.5$, we obtain

\begin{equation}
\frac{\mathrm{MSD}_{\uparrow}(\tau^{\star})}{a^2}
\simeq 3.35,
\qquad
\frac{\mathrm{MSD}_{\downarrow}(\tau^{\star})}{a^2}
\simeq 7.67.
\label{eq:msd_values_tstar}
\end{equation}

The smaller spin-up MSD is consistent with the reduced magnetic-band
dispersion identified in Fig.\ref{fig:spin_bands}. It nevertheless
continues to increase over the displayed interval, demonstrating that
the localized central-site excitation retains a finite overlap with
the residual dispersive spin-up subbands.

To characterize the local spreading law, we evaluate

\begin{equation}
\alpha_{\sigma}^{\mathrm{eff}}(\tau)
=
\frac{
d\ln[\mathrm{MSD}_{\sigma}(\tau)]
}{
d\ln\tau
}.
\label{eq:effective_exponent_results}
\end{equation}

Numerically, the logarithmic derivative is extracted from a local
least-squares fit over 27 neighboring time samples. The interval
$\tau<0.3$ is excluded because both the MSD and its logarithm are
small in the short-time limit. As shown in
Fig.\ref{fig:dynamical_observables}(b), the zero-flux spin-down
component remains close to ballistic propagation, while the
half-flux spin-up component displays a smaller and coherently
oscillating effective exponent. At $\tau^{\star}=4.5$,

\begin{equation}
\alpha_{\uparrow}^{\mathrm{eff}}
\simeq 1.29,
\qquad
\alpha_{\downarrow}^{\mathrm{eff}}
\simeq 1.91.
\label{eq:effective_exponent_values}
\end{equation}

The intermediate spin-up value describes coherent dynamical slowing;
it should not be interpreted as diffusion because the model contains
neither disorder nor a dephasing mechanism.

The participation ratio provides a complementary measure of the
number of appreciably occupied lattice sites. As shown in
Fig.\ref{fig:dynamical_observables}(c), the spin-down component
occupies a substantially larger portion of the lattice than the
spin-up component. At $\tau^{\star}$, the corresponding values are

\begin{equation}
\mathrm{PR}_{\uparrow}(\tau^{\star})
\simeq 30.2,
\qquad
\mathrm{PR}_{\downarrow}(\tau^{\star})
\simeq 64.9.
\label{eq:pr_values_tstar}
\end{equation}

The simultaneous reduction of the MSD and participation ratio shows
that the spin-up component has both a smaller propagation radius and
a more compact spatial distribution.

We further quantify central retention by introducing the finite
region

\begin{equation}
\mathcal R
=
\left\{
i:
|\mathbf r_i-\mathbf r_0|
\leq3a
\right\},
\label{eq:retention_region}
\end{equation}

and use the retention probability defined in
Eq.\eqref{eq:retention_probability}.

Figure\ref{fig:dynamical_observables}(d) shows that the half-flux
spin-up component maintains a larger fraction of its probability
inside $\mathcal R$. At the selected observation time,

\begin{equation}
P_{\mathcal R,\uparrow}(\tau^{\star})
\simeq0.911,
\qquad
P_{\mathcal R,\downarrow}(\tau^{\star})
\simeq0.746.
\label{eq:retention_values_tstar}
\end{equation}

Thus, more than $90\%$ of the spin-up probability remains within
$3a$ of the initial site at $\tau^{\star}$, even though a finite
fraction has leaked through the dispersive magnetic subbands.

For a compact measure of spin selectivity, we define the normalized
MSD contrast as

\begin{equation}
\mathcal C_{\mathrm{MSD}}(\tau)
=
\frac{
\mathrm{MSD}_{\downarrow}(\tau)
-
\mathrm{MSD}_{\uparrow}(\tau)
}{
\mathrm{MSD}_{\downarrow}(\tau)
+
\mathrm{MSD}_{\uparrow}(\tau)
}.
\label{eq:msd_contrast}
\end{equation}

This quantity vanishes when the two spin components spread
identically and becomes positive when the spin-up propagation is
suppressed relative to the spin-down propagation.
Figure\ref{fig:dynamical_observables}(e) shows that the contrast
develops continuously during the evolution and reaches

\begin{equation}
\mathcal C_{\mathrm{MSD}}(\tau^{\star})
\simeq0.392.
\label{eq:msd_contrast_value}
\end{equation}

The resulting contrast is therefore not restricted to an isolated
instantaneous density pattern but persists over an extended
observation interval.

Finally, to verify that the retention contrast is not an artifact of
the particular choice $R=3a$, we calculate the cumulative radial
probability

\begin{equation}
F_{\sigma}(R;\tau^{\star})
=
\frac{
\sum_{
|\mathbf r_i-\mathbf r_0|\leq R
}
n_{i\sigma}(\tau^{\star})
}{
\sum_i n_{i\sigma}(\tau^{\star})
}.
\label{eq:cumulative_radial_probability}
\end{equation}

As shown in Fig.\ref{fig:dynamical_observables}(f), the spin-up
cumulative probability approaches unity at a smaller radius than the
spin-down probability. The two values at $R=3a$ coincide with those
reported in Eq.\eqref{eq:retention_values_tstar}. The radial
dependence therefore confirms that the enhanced spin-up retention is
not tied to a single arbitrarily selected integration radius.

The close agreement among the results for $L=21$, $25$, and $29$
indicates that finite-size effects are negligible over the displayed
time interval. Taken together, the MSD, effective exponent,
participation ratio, retention probability, and cumulative radial
profile establish a pronounced spin-selective dynamical slowing at
the working point. At the same time, the continued growth of the
spin-up MSD and its finite leakage are consistent with coherent
dynamical slowing in a partially flattened spectrum rather than strict
all-state Aharonov--Bohm caging. We next examine whether this dynamical
contrast survives detuning away from the half-flux condition.

\subsection{Flux-controlled reversal of the spin contrast}
\label{subsec:flux_switching}

The identity of the spectrally flattened spin component can be switched
by reversing the spin-dependent part of the synthetic gauge field.
Under

\begin{equation}
\alpha_s\longrightarrow-\alpha_s,
\end{equation}

the effective fluxes are interchanged,

\begin{equation}
\alpha_{\uparrow}
\longleftrightarrow
\alpha_{\downarrow}.
\end{equation}

For example, choosing

\begin{equation}
\alpha_0=\frac{1}{4},
\qquad
\alpha_s=-\frac{1}{4}
\end{equation}

gives

\begin{equation}
\alpha_{\uparrow}=0,
\qquad
\alpha_{\downarrow}=\frac{1}{2}.
\end{equation}

The spin-down component then experiences the half-flux spectrum, while
the spin-up component experiences zero flux. After subtracting the
uniform Zeeman offsets, the two orbital Hamiltonians are interchanged.
For spin-symmetric initial spatial states, the corresponding MSD
contrast therefore changes sign,

\begin{equation}
\mathcal C_{\mathrm{MSD}}
\longrightarrow
-\mathcal C_{\mathrm{MSD}}.
\end{equation}

This controllable reversal distinguishes the mechanism from
spin-independent slowing and demonstrates that the synthetic flux
functions as a reversible dynamical spin switch.

\subsection{Response to flux detuning}
\label{subsec:robustness}
\begin{figure*}[t]
    \centering
    \includegraphics[width=0.95\textwidth]
    {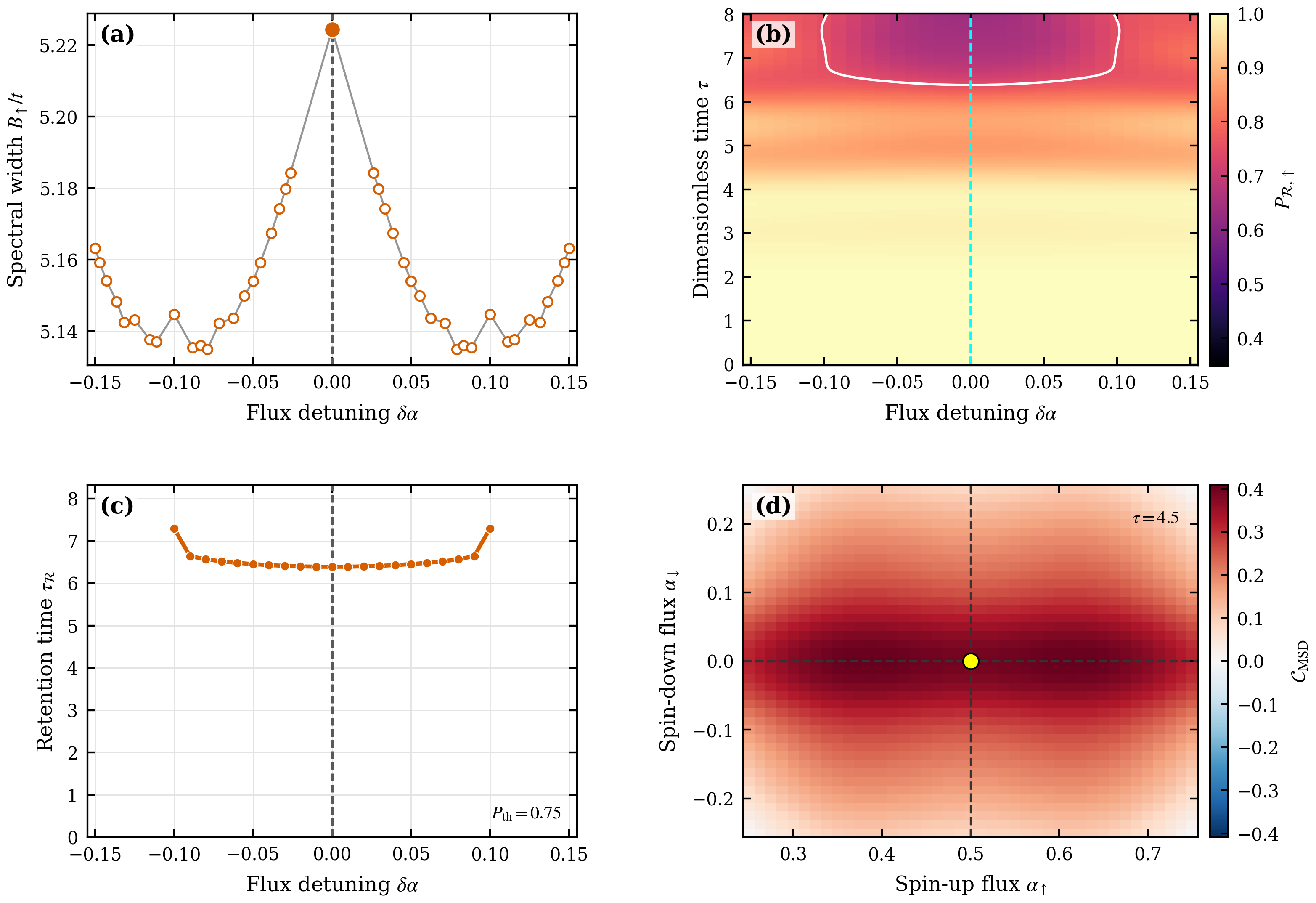}
    \caption{
    Response of the spin-selective spectral and dynamical contrast to
    flux detuning.
    (a) Total spin-up spectral width
    $B_{\uparrow}=E_{\max,\uparrow}-E_{\min,\uparrow}$ as a function of
    $\delta\alpha=\alpha_{\uparrow}-1/2$. The points correspond to
    reduced rational fluxes with denominators $q\leq20$.
    (b) Spin-up finite-region retention probability
    $P_{\mathcal R,\uparrow}(\tau)$ versus evolution time and flux
    detuning. The white contour denotes
    $P_{\mathcal R,\uparrow}=P_{\mathrm{th}}=0.75$, and the cyan dashed
    line marks $\delta\alpha=0$.
    (c) First-passage retention time $\tau_{\mathcal R}$ extracted from
    Eq.\eqref{eq:retention_time_definition}. Upward triangles indicate
    cases for which the threshold is not reached within
    $0\leq\tau\leq8$ and therefore represent the lower bound
    $\tau_{\mathcal R}>8$.
    (d) MSD spin contrast
    $\mathcal C_{\mathrm{MSD}}$ in the
    $(\alpha_{\uparrow},\alpha_{\downarrow})$ plane at $\tau=4.5$.
    Thin solid contours indicate
    $\mathcal C_{\mathrm{MSD}}=0$, while the dashed diagonal denotes the
    flux-exchange path
    $\alpha_{\uparrow}+\alpha_{\downarrow}=1/2$.
    The yellow marker identifies the working point
    $(\alpha_{\uparrow},\alpha_{\downarrow})=(1/2,0)$.
    }
    \label{fig:robustness}
\end{figure*}
We finally examine whether the spin-selective spectral and dynamical
contrast survives deviations from the nominal half-flux working point.
The spin-up flux is written as

\begin{equation}
\alpha_{\uparrow}
=
\frac{1}{2}
+
\delta\alpha,
\label{eq:flux_detuning}
\end{equation}

where $\delta\alpha$ denotes the flux detuning.

Care is required when comparing magnetic-subband bandwidths at
different rational fluxes. Because the number of magnetic subbands
depends on the denominator $q$, the mean subband bandwidth
$\overline{W}_{\sigma}$ is affected by magnetic-band folding and is
therefore not directly comparable between arbitrary values of $q$.
For the detuning analysis, we instead use the total orbital spectral
width

\begin{equation}
B_{\uparrow}(\alpha_{\uparrow})
=
\max_{n,\bm{k}}
E_{n\uparrow}(\bm{k})
-
\min_{n,\bm{k}}
E_{n\uparrow}(\bm{k}),
\label{eq:total_spectral_width}
\end{equation}

which is independent of the choice of magnetic unit cell.

To characterize the dynamical response, we define the first-passage
retention time $\tau_{\mathcal R}$ through

\begin{equation}
P_{\mathcal R,\uparrow}
\left(
\tau_{\mathcal R}
\right)
=
P_{\mathrm{th}},
\label{eq:retention_time_definition}
\end{equation}

where $P_{\mathrm{th}}=0.75$. The region $\mathcal R$ is a circle of
radius $3a$ centered on the initially populated $C$ site and is held
fixed throughout the comparison. If the retention probability does not
fall below the threshold within the observation interval
$0\leq\tau\leq8$, only the lower bound
$\tau_{\mathcal R}>8$ is reported.

Figure\ref{fig:robustness}(a) shows that the total spectral width
changes continuously around half flux. At the nominal working point,

\begin{equation}
B_{\uparrow}(1/2)
=
5.224\,t.
\label{eq:half_flux_spectral_width}
\end{equation}

For $|\delta\alpha|\simeq0.10$, the total spectral width decreases
only modestly to
$B_{\uparrow}\simeq5.145\,t$, approximately $1.5\%$ below its value
$5.224\,t$ at $\alpha_{\uparrow}=1/2$. Thus,
$B_{\uparrow}$ varies smoothly across the half-flux point and exhibits
no singular feature there. We emphasize, however, that the total
spectral width alone is not a criterion for exact caging, because a
set of energetically separated flat bands can retain a finite overall
spectral span. The decisive spectral evidence is instead
band-resolved: four of the six magnetic subbands at
$\alpha_{\uparrow}=1/2$ have finite bandwidths, as shown in
Fig.\ref{fig:spin_bands}(b), thereby excluding an all-bands-flat
Aharonov--Bohm cage.

The corresponding real-time response is shown in
Fig.\ref{fig:robustness}(b). Over the time window relevant to
Fig.\ref{fig:density_dynamics}, the retention probability depends
only weakly on flux detuning. At $\tau=4.5$,
$P_{\mathcal R,\uparrow}$ remains between $0.910$ and $0.911$ throughout
the interval $|\delta\alpha|\leq0.15$. The enhanced central retention
is therefore robust against moderate flux imperfections over this
finite observation window. At the same time, the absence of a sharp
feature at $\alpha_{\uparrow}=1/2$ confirms that the observed response
is a robust finite-time dynamical slowing rather than localization at
a singular exact-caging point.

Differences become more apparent at longer evolution times. At
$\delta\alpha=0$, the threshold
$P_{\mathcal R,\uparrow}=0.75$ is first reached at

\begin{equation}
\tau_{\mathcal R}(0)
=
6.388.
\label{eq:half_flux_retention_time}
\end{equation}

As shown in Fig.\ref{fig:robustness}(c), the first-passage time
increases to approximately $7.295$ at
$|\delta\alpha|=0.10$. For
$|\delta\alpha|\geq0.11$, the probability remains above the selected
threshold throughout the simulated interval, giving
$\tau_{\mathcal R}>8$.

This nonmonotonic behavior is important. For the central-$C$-site
initial state considered here, half flux is not the point of maximum
retention time. The detuning dependence instead results from coherent
redistribution among the residual dispersive magnetic subbands.
Consequently, $\tau_{\mathcal R}$ is a state- and threshold-dependent
finite-time diagnostic rather than a universal cage lifetime.

Figure\ref{fig:robustness}(d) demonstrates that the dynamical spin
contrast remains substantial over a finite region of the two-flux
parameter space. At the working point,

\begin{equation}
\mathcal C_{\mathrm{MSD}}
\left(
\frac{1}{2},0;\tau=4.5
\right)
=
0.392.
\label{eq:working_point_contrast}
\end{equation}

Interchanging the two effective fluxes reverses the sign of the
contrast,

\begin{equation}
\mathcal C_{\mathrm{MSD}}
\left(
\alpha_{\uparrow},
\alpha_{\downarrow}
\right)
=
-
\mathcal C_{\mathrm{MSD}}
\left(
\alpha_{\downarrow},
\alpha_{\uparrow}
\right),
\label{eq:contrast_exchange_symmetry}
\end{equation}

so that the reverse working point
$(\alpha_{\uparrow},\alpha_{\downarrow})=(0,1/2)$ produces the opposite
spin separation. Along the dashed flux-exchange path, the contrast
passes continuously through zero at
$\alpha_{\uparrow}=\alpha_{\downarrow}=1/4$.

These results establish robustness in the experimentally relevant
finite-time sense: moderate flux detuning does not eliminate the
spin-resolved spatial contrast. At the same time, neither the spectral
width nor the retention time exhibits the singular behavior expected
for an exact Aharonov--Bohm cage. Figure\ref{fig:robustness} therefore
reinforces the interpretation of the present effect as controllable,
spin-selective dynamical slowing rather than strict all-state caging.
\section{Conclusion}
\label{sec:conclusion}

We have shown that a spin-dependent synthetic flux converts
Lieb-lattice interference into a reversible internal-state-resolved
transport contrast. At
$\alpha_0=\alpha_s=1/4$, the spin-up and spin-down components experience
half and zero flux, respectively, while evolving under otherwise
identical nearest-neighbor Hamiltonians. Because a uniform longitudinal
Zeeman term changes only the overall phase of each conserved spin
sector, the observed difference is an orbital effect of the
spin-dependent Peierls phases.

At $\alpha_{\uparrow}=1/2$, the doubled magnetic unit cell contains
two zero-energy flat subbands associated with the
sublattice-imbalance flat-band sector, whereas the other four
magnetic subbands remain dispersive. The half-flux spectrum therefore
does not exhibit an all-band collapse.
The
central-$C$-site wave packet consequently continues to expand. At
$\tau=4.5$, its spin-up MSD is $3.348a^2$, compared with $7.673a^2$ for
spin down, and the corresponding probabilities retained within radius
$3a$ are $0.911$ and $0.746$. The continued growth of the spin-up MSD
and its finite leakage confirm that the result is coherent finite-time
slowing in a partially flattened spectrum, not strict caging of an
arbitrary localized state.

The mechanism remains controllable beyond this working point. Reversing
$\alpha_s$ interchanges the two orbital Hamiltonians and reverses the
sign of the MSD contrast. Moderate flux detuning preserves a substantial
spatial separation over the pre-reflection observation window, although
the state- and threshold-dependent retention time is not maximized
exactly at half flux. This behavior further distinguishes the observed
contrast from a singular all-bands-flat cage.

From an atomic, molecular, and optical physics perspective, the protocol
requires state-dependent tunneling phases, internal-state preparation,
and spin-resolved density imaging---capabilities already associated with
optical-lattice gauge-field experiments. The resulting framework
provides both a reversible method for internal-state-selective
matter-wave routing and a practical spectral-to-real-space test for
distinguishing exact Aharonov--Bohm caging from finite-time dynamical
slowing. The same criteria apply directly to photonic flat-band
simulators, where propagation distance plays the role of evolution time.

\section*{Data Availability}

The numerical data and source code supporting the findings of this study are available from the corresponding author upon reasonable request.

\begin{acknowledgments}
This work was supported by the National Natural Science Foundation of China (Grants no. 92365208), National Key Research and Development Program of China (Grants no. 2021YFA0718300 and no. 2021YFA1400900), the National Key Research and Development Program of China (No.~2023YFB2407900), National Key Research and Development Program of China 
(No.~2024YFE0200502).
\end{acknowledgments}


\bibliography{Refs}

\end{document}